\documentclass[12pt]{article}

\begin{document}

\title{Nonlinear gauge interactions - A solution to the ``measurement
problem" in quantum mechanics?}

\author{Johan Hansson
\\ \textit{Department of Physics} \\ \textit{University of G\"{a}vle}
 \\ \textit{SE-801 76 G\"{a}vle, Sweden}}

\maketitle

\begin{abstract}
We propose that the mechanism responsible for the ``collapse of
the wave function" (or ``decoherence" in its broadest meaning) in
quantum mechanics is the nonlinearities already present in the
theory via nonabelian gauge interactions. Unlike all other models
of spontaneous collapse, our proposal is, to the best of our
knowledge, the only one which does not introduce any new elements
into the theory. Indeed, unless the gauge interaction
nonlinearities are not used for exactly this purpose, one must
then explain why the violation of the superposition principle
which they introduce does not destroy quantum mechanics. A
possible experimental test of the model would be to compare the
coherence lengths for, \textit{e.g.}, electrons and photons in a
double-slit experiment. The electrons should have a finite
coherence length, while photons should have a much longer (in
principle infinite) coherence length.
\\
\\
PACS numbers: 03.65.-w, 03.65.Bz, 11.15.-q
\end{abstract}


We start by noting an apparent paradox in the presently most
fundamental description of (experimentally tested) physical
reality.
\begin{itemize}
  \item 1) The absolute backbone of quantum mechanics is the
  \textit{superposition principle} \cite{Dirac} (\textit{e.g.}, interfering
  amplitudes, summation of Feynman diagrams, etc). It is also
  well known that superposition requires \textit{linear} equations
  (\textit{i.e.}, the sum of two different solutions to a nonlinear
  equation is generally \textit{not} a solution, ruining superposition).
  The Hilbert space of quantum mechanics and the Fock space of
  quantum field theory are \textit{linear} spaces (based on the
  superposition requirement), suitable for linear mappings or
  operators.
  \item 2) Nonabelian gauge field theories describing the fundamental
  interactions obey \textit{nonlinear}
  evolution equations in the gauge fields. This is in apparent
  contradiction to point 1). For convenience, we write down the
  evolution equations for pure Yang-Mills fields below. Although
  the fermion evolution obeys linear equations, they become ``contaminated"
  by nonlinearities through the interaction.
\end{itemize}

The nonabelian vector gauge fields are governed by a set of
coupled, second order, nonlinear PDEs on Minkowski spacetime. (The
general argument for gravity is the same, but involve tensor
fields on a dynamical spacetime. We do not explicitly write down
those equations.) For pure Yang-Mills the evolution equations are
given by the following formula,
\begin{equation}
(\partial^{\mu} - g [A^{\mu}, \, ])_a ^b (\partial_{\mu} A_{\nu} -
\partial_{\nu} A_{\mu} -g[A_{\mu}, A_{\nu}])_b = 0 ,
\end{equation}
where $g$ is the coupling constant and $a$, $b$ are indices of the
gauge group (\textit{i.e.}, $a, b \in 1,2,3$ for $SU(2)$ and $a, b
\in 1,...,8$ for $SU(3)$). Summation over repeated indices is
implied. The operator (``covariant derivative") at the left works
according to $(\partial^{\mu} - g [A^{\mu}, \, ])(anything) =
\partial^{\mu}(anything)- g [A^{\mu}, anything]$. We see that we
get highly nonlinear (quadratic and cubic) terms in the gauge
fields, especially when the coupling constant, $g$, is large. The
commutator terms (square brackets) vanish identically for abelian
fields (\textit{e.g.} photons) because the gauge fields then
commute, leaving only the ordinary, linear Maxwell equations.

 In the Feynman path-integral formulation of quantum
mechanics \cite{Feynman} the nonlinearities can be ``hidden" in
the action functional, but as the Schr\"{o}dinger, Heisenberg and
path-integral formulations are equivalent, a problem in one of
them must translate into a problem in all formulations.

Instead of trying to reconcile the two apparently contradictory
statements above (by, for instance, modifying the rules of quantum
mechanics), introduced by nonabelian gauge fields, we instead
propose to turn a vice into a virtue by postulating that it is the
dynamical nonlinear interaction terms which break the
superposition of different quantum states of a system,
\textit{i.e.} acting as the physical mechanism which reduces the
state vector, or ``collapses the wave function" in the less
general Schr\"{o}dinger setting. We thus get a self-induced
collapse - ``SIC", into the ordinary world of chairs, tables,
people and indeed also recorded elementary particle tracks in a
photographic emulsion, a bubble chamber, or a modern multi-purpose
computer-aided detector.

That a quantum mechanical state must be able to ``self-decohere"
is imperative in quantum cosmology, the quantum mechanical
treatment of the whole universe, where no ``outside" observer
exists. The self-induced collapse puts an end to the infinite
regress of quantum superposition, where first the measuring
apparatus obtains a quantum mechanical nature, then the observer,
and so on, ad infinitum, until the whole universe consists of
infinitely many superimposed quantum states, without any one of
them actually being ``realized". The ``many worlds" interpretation
of Everett \cite{Everett} purports to solve this problem by
assuming that we only see events which take place in one of these
branching universes, but it seems that the fundamental question of
\textit{when}, and how, the universe actually branches is
unanswered by that model (this being the equivalent of the
``measurement problem" in the orthodox interpretation).

We now turn to the actual implementation of our idea of
self-induced collapse. For simplicity, we choose the following
(non-covariant) expression for the (average) self-decoherence
time.

\begin{equation}
\tau = \frac{\hbar}{E_{N.L.}},
\end{equation}
where $E_{N.L.}$ is the energy stored in the nonlinear field
configuration of the nonabelian interaction (which in turn depends
on the strength of the coupling). Observe that the relation is
\textit{not} an uncertainty relation, despite its identical form,
as $\tau$ and $E_{N.L.}$ are not uncertainties. As we want the
energy of the full nonlinear theory, and we cannot today
explicitly calculate this inherently non-perturbative quantity, we
take the energy to be a characteristic energy for the interaction.
If, for instance, for QCD, we as a rough approximation take the
energy to be $E_{N.L.} \sim \Lambda_{QCD} \sim 0.2$ GeV, we get as
a rough ``ballpark" figure $\tau_{QCD} \sim 10^{-23}$s for the
decoherence time for \textit{strong} QCD (\textit{e.g.}, inside a
non-disturbed hadron). Although the exact result probably will
differ by many orders of magnitude, this may explain why
(semi-)classical models work so well for strong QCD, as the
stronger the interaction is, the more ``classical" it behaves
according to our mechanism. In QCD the energy stored in gauge
fields decreases as the absolute energy of the interaction
increases, due to asymptotic freedom.

In our model, the \textit{fields} are the fundamental entities
which obey quantum mechanics, the particle aspect appearing each
time a self-collapse takes place. That is, the quantum mechanical
(linear, unitary) evolution is constantly punctuated by (possibly
random) ``hits" of self-collapse at an average frequency of
$\tau^{-1}$. This is similar to the case in orthodox quantum
mechanics where an observation (or the initial preparation of a
state) suddenly ``realizes" one of the potential outcomes, after
which the unitary (linear) evolution of the state takes over until
the next observation. A ``macroscopic" piece of matter has such a
high energy stored in nonlinear field configurations that $\tau =
\frac{\hbar}{E_{N.L.}} \sim 0$, approximating a continuously
collapsing state, \textit{i.e.}, a classical state. The model thus
forbids quantum mechanical effects to ``invade" the macroscopic
world, and hence resolves the ``Schr\"{o}dinger's cat" paradox
\cite{Schrodinger} and related questions such as ``Wigner's
friend" \cite{Wigner}, etc.

Note that \textit{any} significant nonlinear interaction, whether
as part of a ``measurement" carried out by conscious beings
\cite{Schrodinger,Wigner} or not, bring about the decoherence of
interfering amplitudes into (semi-)classical states. Conscious
observation is therefore only a \textit{special case} of the more
general nonlinearity, as all ``measuring apparatuses" (including
human beings!) consist of both weakly (all particles) and strongly
(quarks) nonlinearly interacting constituents. Hence, there should
be no need to introduce the \textit{mind} into the interpretation
of quantum mechanics at a fundamental level.

For pure QED the nonlinear terms are absent, hence a hypothetical
world built by QED alone would never be classical. It also
explains why, \textit{e.g.}, atomic physics works so closely to
orthodox quantum mechanics, as it is being ``classicalized" only
by (very) weak interaction effects. Were it not for the existence
of the other interactions besides QED, we would indeed have
quantum mechanical superpositions of whole universes,
\textit{i.e.}, the ``many worlds" interpretation of quantum
mechanics by Everett \cite{Everett}.

The difference between our proposal for self-induced collapse, and
other models aiming at the same goal, is that, as far as we know,
all other models postulate additional equations and/or variables,
\begin{itemize}
  \item Decohering histories \cite{Griffiths,GellMann}: new fundamental
 principle of irreversible coarse graining + additional
constraints to remove ``too many" decoherent histories
  \item Altered Schr\"{o}dinger equation: obvious extra
  (non-unitary\cite{GRW} or nonlinear\cite{Pearle}) term in Schr\"{o}dinger equation
  \item Bohm QM \cite{Bohm}: additional (nonlinear) evolution equation
for objective positions
\end{itemize}
whereas we use only nonlinearities which are \textit{already
present} in the dynamics of the accepted standard model of
particle physics. Another difference is that, to our knowledge,
all other models for spontaneous collapse are non-relativistic,
whereas our scheme is based on covariant theories. Also, it must
be stressed that if the nonlinearities introduced by the
nonabelian gauge fields are \textit{not} used to explain the
decoherence to (quasi)classical behaviour, it must instead be
explained how they can be reconciled with the superposition
principle of quantum mechanics.

It is well known that a nonlinear mapping is non-reversible, as
there in that case does not exist a unique inverse mapping
(\textit{i.e.}, the mapping is not one-to-one). We therefore
propose that the nonlinear gauge interaction is the physical
``mechanism" of the ``irreversible amplification" emphasized by
Bohr as being necessary to produce classical, observable results
from the quantum mechanical formalism. Even though Bohr himself
denounced the need, or even the possibility, to give a physical
description of this ``mechanism" \cite{Bohr}, we believe that the
central problem for truly understanding quantum mechanics lies in
the quantum measurement problem. For instance, it is \textit{only}
there, in the collapse of the wave function, that the
undeterminacy of quantum mechanics enters. It may even be
possible, if not entirely likely, that deterministic chaos in the
nonlinear self-interaction can be responsible for the seemingly
statistical character of quantum mechanics.

Our model can be experimentally tested, at least in principle, as
differently charged (electric, weak isospin, color,...) particles
should have different coherence lengths. In a double-slit
experiment, for instance, the photon should have a much longer (in
principle infinite) coherence length than, \textit{e.g.},
electrons which ought to have a finite coherence length due to
nonlinear weak interactions. As the full nonlinear calculations
are very complicated, it is not possible to quantitatively predict
the coherence lengths at the present time, but if it turns out
that electrons experimentally have shorter coherence lengths than
photons it would strengthen our hypothesis.

Our model could also have importance for (the not yet existing
theory of) quantum gravity. Weak gravity would have extremely long
decoherence times, completely swamped by the other interactions.
However, exactly where quantum gravity is expected to become
important (\textit{i.e.}, at the Planck mass/energy scale), we see
that it spontaneously decoheres. Hence a strong \textit{quantum}
gravity might not exist in this scheme.

The collapse postulated in orthodox quantum mechanics is not
relativistically covariant, as it is instantaneous (over all
space), which is not a covariant concept. Only the (deterministic)
unitary development of the state is taken into account by the
relativistic Dirac equation and, more generally, by quantum field
theory. As our scheme for collapse is based on covariant
gauge-field theories, it might be possible to describe the
collapse of the state in a covariant way, although our present
attempt, which for \textit{simplicity} singles out the energy
accumulated in the nonlinear field configurations, is not
covariant. On the other hand, it might be that the collapse should
be described by an inherently non-local mechanism, as it seems
that quantum mechanics at its very foundation \textit{is}
non-local, as given by the results of Aspect \textit{et al.}
\cite{Aspect}, and more recent experiments on quantum
non-separability. As a nonlinear theory also in some cases can be
non-local, it would be interesting to investigate if this model of
spontaneous collapse can account for such non-local effects. Work
in this direction is in progress \cite{HanssonDugne}, together
with a detailed investigation of the nonlinear terms in the
nonabelian evolution equations, as a means to better understand
the quantitative details of the proposed mechanism for
self-collapse.

In conclusion, we have emphasized that automatic dynamical
collapse of the wave function in quantum mechanics may already be
implicit in the existing dynamical theory of nonabelian
(\textit{i.e.}, nonlinear) gauge fields. These include the weak
interaction, QCD, gravity, and any other nonabelian fields which
eventually may be found in the future. The nonlinear
self-interaction terms break the fundamental superposition
principle of quantum mechanics, hence making it plausible that
they can be just the right physical mechanism for the purpose of
collapse.

\end{document}